\documentclass[12pt]{iopart}

\usepackage{subfigure}
\usepackage{rotating}

\newcommand{\msun}{M_\odot}
\newcommand{\be}{\begin{equation}}
\newcommand{\ee}{\end{equation}}
\newcommand{\bea}{\begin{eqnarray}}
\newcommand{\eea}{\end{eqnarray}}
\newcommand\flow{f_{\rm low}}

\newcommand\fmerg{f_{\rm merg}}
\newcommand\fring{f_{\rm ring}}
\newcommand\fcut{f_{\rm cut}}

\newcommand\Mcrit{M_{\rm crit}}
\newcommand{\prl}{{\it Phys. Rev. Lett. }}
\newcommand{\prd}{{\it Phys. Rev.} D }
\newcommand{\cqg}{{\it Class. Quantum Grav. }}

\begin{document}

\title[]{Testing the validity of the phenomenological gravitational waveform models for nonspinning binary black hole searches at low masses}

\author{Hee-Suk Cho}
\ead{chohs1439@pusan.ac.kr}
\address{Korea Institute of Science and Technology Information, Daejeon 305-806, Korea}

\begin{abstract}
The phenomenological gravitational waveform models, which we refer to as PhenomA, PhenomB and PhenomC, generate full inspiral-merger-ringdown waveforms of coalescing binary back holes (BBHs). These models are defined in the Fourier domain,  thus can be used for fast matched filtering in the gravitational wave search. 
PhenomA has been developed for nonspinning BBH waveforms, while PhenomB and PhenomC were designed to model the waveforms of BBH systems with nonprecessing (aligned) spins, but can also be used for nonspinning systems.
In this work, we study the validity of the phenomenological models for nonspinning BBH searches at low masses, $m_{1,2}\geq 4 \msun$ and $m_1+m_2\equiv M \leq 30 \msun$, with Advanced LIGO. 
As our complete signal waveform model, we adopt EOBNRv2 that is a time-domain inspiral-merger-ringdown waveform model. 
To investigate the search efficiency of the phenomenological template models,
we calculate fitting factors by exploring overlap surfaces.
We find that only PhenomC is valid to obtain the fitting factors better than 0.97 in the mass range of $M<15 \msun$.
Above $15 \msun$, PhenomA is  most efficient in symmetric mass region, 
PhenomB is  most efficient in highly asymmetric mass region, and
PhenomC is  most efficient in the intermediate region.
Specifically, we propose an effective phenomenological template family that can be constructed by employing the phenomenological models in four subregions individually.
We find that fitting factors of the effective templates are better than 0.97 in our entire mass region and mostly greater than 0.99.
\end{abstract}



\section{Introduction}
Coalescing binary black holes (BBHs) are among the most promising sources 
of gravitational wave (GW) transients for ground-based detectors, such as LIGO~\cite{Aas15} and Virgo~\cite{Ace15}.
GW signals emitted from BBHs are conventionally divided into three phases:
inspiral, merger and ringdown (IMR).
Inspiral waveforms can be accurately obtained by using post-Newtonian (PN)  
approximations. When the system reaches the 
ultra-relativistic regime, however, merger-ringdown waveforms should be calculated from numerical relativity (NR) simulations.
A full IMR waveform can be constructed by combining a PN inspiral waveform and a NR merger-ringdown waveform.
However, obtaining merger-ringdown waveforms from NR simulations is computationally very expensive.
Thus, inspiral waveforms have mainly been used for low mass binaries below $\sim 30 \msun$ in ground-based GW data analyses~\cite{Aba12}.
It has been anticipated that 
the inspiral waveforms would yield accurate analysis results  for low mass binaries 
because in those systems the inspiral phase is likely to have most of the signal power in detection frequency band of  ground-based detectors.

However, several authors have pointed out that
the absence of merger-ringdown phase can decrease the search efficiency, thus induce a non-negligible loss in detection rate even for  low mass systems.
When using inspiral-only template waveforms, a loss in signal-to-noise ratio (SNR) tends to increase with increasing total mass ($M$) of the signal, consequently the detection rate  becomes reduced below $90\%$ of that achievable by an IMR search if $M$ exceeds some critical value $\Mcrit$ (this is explained in more detail at the end of section~\ref{sec3.1}).
Farr \etal~\cite{Far09} obtained the values of $\Mcrit$ about $10 - 15 \msun$ using IMR signals generated from EOBNR model  and TaylorF2 templates with the initial LIGO sensitivity.
A similar study has been carried out by Brown \etal~\cite{Bro13} for Advanced LIGO, and they found $\Mcrit \sim 11.4 \msun$.
Buonanno \etal~\cite{Buo09} also showed that $\Mcrit \sim 12 \msun$ for various PN inspiral templates
with EOBNR signals.
Assuming phenomenological IMR signals, Ajith~\cite{Aji08b} and Cho~\cite{Cho15} found that $\Mcrit \sim 15 \msun$ for inspiral templates.
While the former used $3.5$PN TaylorT1 templates, in the latter the inspiral waveforms were constructed by taking the inspiral parts into account from the original phenomenological IMR waveforms
to avoid the systematic effect that arises when the model for signals is different from the model for templates.

Efforts to establish the analytic IMR waveform models have been made over the past years.
Those works have been carried out by constructing phenomenological families of waveforms in the Fourier domain by using  PN-NR hybrid waveforms.
The first phenomenological family (PhenomA), which can model the IMR waveforms of nonspinning BBHs, has been proposed by Ajith \etal~\cite{Aji08b,Aji07a,Aji08a}.
The second family (PhenomB) has been developed by Ajith \etal~\cite{Aji11} by extending PhenomA to the case of nonprecessing spinning BBHs.
The third  family (PhenomC) has been made by Sanataria \etal~\cite{San10} also for nonprecessing spinning BBHs.
A recent phenomenological family (PhenomP) allows us to have IMR waveforms of  precessing BBHs \cite{Han13}.
On the other hand,  time-domain IMR waveform models have also been developed using the effective-one-body (EOB) approach~\cite{Buo99,Buo00}.
Accuracies of EOB models have been improved by calibrating them to NR waveforms. 
The nonspinning EOBNR model described in Ref.~\cite{Buo07} has been
used in LIGO and Virgo to search for high mass BBHs for the first time~\cite{Aba11}\footnote{In the observational analysis of Ref.~\cite{Aba11},
the sensitivity of a search with nonspinning EOBNR templates was estimated using simlulated PhenomA signals, which is roughly the opposite of our approach.},
and this model has been continuously upgraded  as new NR simulations have been produced.
The state-of-art calibrations of this model can be found in Refs.~\cite{Pan11,Dam13}.
A nonprecessing spinning EOBNR model has also been proposed by Taracchini \etal~\cite{Tar12} and further calibrated~\cite{Tar14} to 
38 NR waveforms produced by the SXS Collaboration~\cite{SXS}. 
Recently, Pan \etal~\cite{Pan14a} have described a general procedure to generate precessing EOBNR waveforms starting from a nonprecessing spinning EOBNR model given in Ref.~\cite{Tar12}, and 
they found remarkable agreement with two precessing NR waveforms.
The EOBNR models of Refs.~\cite{Pan11,Tar12,Tar14} have been  implemented in the  LSC Algorithm Library (LAL)~\cite{lal} under the names of EOBNRv2, SEOBNRv1 and SEOBNRv2.

The stability of nonspinning EOBNR model has been investigated by Pan \etal~\cite{Pan14b}.
They showed that those waveforms of any length are sufficiently accurate for data analysis with advanced GW detectors.
Recently, in addition, that has been anecdotally confirmed by Szil\'agyi \etal~\cite{Szi15} by using a 
long NR simulation in which the gravitational waveform was long enough to cover the entire frequency band of advanced GW detectors
for a nonspinning binary at mass ratio $m_1/m_2=7$ with a total mass as low as $M=45.5 \msun$.
They also found that existing phenomenological IMR waveforms display much greater disagreement with the NR simulation.
EOBNRv2 is now believed to be sufficiently accurate to search for nonspinning BBHs with Advanced LIGO.

However,  EOBNRv2 waveforms are much slower to generate than the phenomenological models since a complicated system
of ordinary differential equations has to be solved over a long time interval with small time steps.
Moreover, time-domain waveforms need to be Fourier-transformed to perform the match, which is defined as a frequency-domain inner product weighted by the detector noise.
We found that the speed of generating Fourier-transforming EOBNRv2 waveforms can be two orders of magnitude slower than the speed of generating phenomenological waveforms.
Thus, the phenomenological waveform families can be used
to densely cover the parameter space, particularly at low masses.  
However, the use of the phenomenological
waveforms may result in a loss in search efficiency due to
inaccuracies of the approximations.

In this work, we adopt as our complete signal waveform model EOBNRv2 and 
test the validity of the phenomenological template models in searches for nonspinning BBHs with Advanced LIGO.
We take into account all the spherical harmonics modes of the EOBNRv2 waveforms available in the LAL, 
those are the leading (2, 2) mode  and  the four sub-leading modes (2, 1), (3, 3), (4, 4) and (5 ,5).
Because the phenomenological fitting coefficients of PhenomB are defined differently from those of PhenomA,
a PhenomB waveform with masses ($m_1, m_2$) and a zero spin is different from a PhenomA waveform with the same masses.
A PhenomP waveform with masses ($m_1, m_2$) and a zero spin is the same as a PhenomC waveform with the same masses and a zero spin 
because PhenomP is derived from PhenomC just by adding an effective precession spin.
Therefore, we  only consider PhenomA, PhenomB and PhenomC models for our nonspinning BBHs.
When using PhenomB and PhenomC as our templates, we will neglect the spin effect by choosing zero spins in the wave functions.
In this work, we focus on low mass systems with $m_{1,2}\geq 4 \msun$ and $m_1+m_2\equiv M \leq 30 \msun$, 
and investigate the fitting factors for each template model.
In particular, we show that for nonspinning BBH searches a template bank made up of several models is feasible 
and can be much more efficient.
For this purpose, we propose an effective template family that is made up of the three phenomenological models and 
show that this template family can have the fitting factors better than 0.97 for all signals in our entire mass region.


\section{Phenomenological waveforms}\label{sec2}
Obtaining phenomenological models involves finding fitting parameters of the analytic waveform family using PN-NR hybrid waveforms. 
The phenomenological waveform families are defined  in the Fourier domain as the  form,
\be
{\tilde h}_{\rm phen}(f) = {A}_{\rm eff}(f) \, e^{\Psi_{\rm eff}(f)},
\label{eq.phenomA}
\ee
where ${A}_{\rm eff}(f)$ and $\Psi_{\rm eff}(f)$ are the effective amplitude and the effective phase, and those are modeled separately.
The model waveforms are parameterised by their physical parameters such as total mass $M$, symmetric mass ratio $\eta \equiv m_1m_2/M^2$ and effective spin parameter $\chi\equiv (1+\delta)\chi_1/2+(1-\delta)\chi_2/2$ where $\delta \equiv (m_1-m_2)/M$ and $\chi_i \equiv S_i/m_i^2$, $S_i$ being the spin angular momentum of the {\it i}th BH.
In this section, we only provide the functional forms of the phenomenological  models.
More details can be found in \cite{Aji08b,Aji07a,Aji08a}, \cite{Aji11} and \cite{San10}  for PhenomA, PhenomB and PhenomC, respectively.


\subsection{PhenomA}
Amplitude spectrum of a PhenomA waveform  is divided by two phenomenological frequencies, $\fmerg$ and $\fring$,
and terminates at the cutoff frequency $\fcut$:
\bea \label{eq.Aeff}
{A_{\rm eff}} = A \fmerg^{-7/6} 
 \left\{ \begin{array}{ll}
\left(f/\fmerg\right)^{-7/6}    &\textrm{if $ f < \fmerg$}   \\
\left(f/\fmerg\right)^{-2/3}    &\textrm{if $\fmerg \leq f < \fring$} \\
w \, {\cal L}(f,\fring,\bar{\sigma})  &\textrm{if $\fring \leq f \leq \fcut$,}\\
\end{array} \right.
\eea
where $A$ is the wave amplitude factor whose value depends on the binary masses and five extrinsic parameters determined by
the sky location and the binary orientation. ${\cal L}(f,\fring,\bar{\sigma}) \equiv \left(\frac{1}{2 \pi}\right) 
\frac{\bar{\sigma}}{(f-\fring)^2+\bar{\sigma}^2/4}$
is a Lorentzian function that has a  width $\bar{\sigma}$, and that is centered
around the frequency $\fring$. The normalization constant,
$w \equiv \frac{\pi \bar{\sigma}}{2} \left(\frac{f_{\rm ring}}
{f_{\rm merg}}\right)^{-2/3}$, is chosen so as to make 
${A}_{\rm eff}(f)$ continuous across the transition frequency $\fring$. 
The parameter $\fmerg$ is the frequency at which the power-law changes 
from $f^{-7/6}$ to $f^{-2/3}$. 
The phenomenological parameters $\fmerg, \fring, \bar{\sigma}$ and $\fcut$ 
can be obtained by fitting their formulas to the PN-NR hybrid waveforms, and those 
are finally given in terms of  $M$ and  $\eta$. The effective phase is expressed as
\be
\Psi_{\rm eff}(f) = 2 \pi f t_c + \phi_c + \frac{1}{\eta}\,\sum_{k=0}^{7}  (x_k\,\eta^2 + y_k\,\eta + z_k)   (\pi M f)^{(k-5)/3},
\label{eq.effectivephase}
\ee
where $t_c$ and $\phi_c$ are the coalescence time and the coalescence phase.
The coefficients  introduced in the phenomenological parameters and the effective phase  are 
tabulated in Table 1 of \cite{Aji08b}. 
The accuracy of PhenomA templates was examined by using ($\sim30$) PN-NR hybrid waveforms finely spaced in the parameter range $m_1/m_2\equiv q \leq 4$ and $50 \leq M/\msun \leq 200$,
in which the fitting factors were found to be greater than 0.99 for Initial LIGO, Advanced LIGO and Virgo.


\subsection{PhenomB}
PhenomB corresponds to an extended version of PhenomA to  nonprecessing spinning BBHs
by incorporating the single spin parameter $\chi$.
The effective amplitude is defined by
\bea \label{eq.Aeff}
{A_{\rm eff}} = A f_1^{-7/6}
 \left\{ \begin{array}{ll}
\left(f/f_1\right)^{-7/6} (1+\Sigma^3_{i=2}\alpha_iv^i)   &\textrm{if $f < f_1$}   \\
w_m \left(f/f_1\right)^{-2/3} (1+\Sigma^2_{i=1}\epsilon_iv^i)   &\textrm{if $f_1 \leq f < f_2$} \\
w_r \, {\cal L}(f,f_2,\bar{\sigma})  &\textrm{if $f_2 \leq f \leq \fcut$},\\
\end{array} \right.
\eea
where $v\equiv (\pi M f)^{1/3}$, $\alpha_{2,3}$ are given in terms of $\eta$ and $\chi$, $\epsilon_{1,2}$ are given in terms of $\chi$ and
the normalization constants $w_m$ and $w_r$ are  chosen so as to make ${A}_{\rm eff}(f)$ continuous across the transition frequency $f_2$ and $f_1$, respectively.
The effective phase is defined by
\be
\Psi_{\rm eff}(f) = 2 \pi f t_c + \phi_c + \frac{2}{128\eta v^5} \left(1+\sum_{k=2}^{7} \psi_k v^k \right).
\label{eq.effectivephase}
\ee
The coefficients  introduced in this model are tabulated in table 1 of \cite{Aji11}. 
PhenomB templates were examined in the parameter range $q \leq 10$, and $ M \leq 400\msun$,
in which the fitting factors were greater than 0.965 for Initial LIGO.

\subsection{PhenomC}
PhenomC has also been developed for nonprecessing spinning BBHs.
The wave amplitude terminates at $\fcut=0.15/M$, and that  is constructed from  two parts as
\be
A_{\rm eff}=A_{\rm PM}(f)w^-_{f_0}+A_{\rm RD}(f)w^+_{f_0},
\ee
where $A_{\rm PM}$ is the premerger amplitude calculated by a PN inspiral amplitude with the addition of a higher order frequency term:
\bea
A_{\rm PM}(f)=A_{\rm PN}(f)+\gamma_1f^{5/6}, \\
A_{\rm PN}=C\Omega^{-7/6}\left(1+\sum^5_{k=2}\gamma_k\Omega^{k/3} \right),
\eea
where $\Omega=\pi M f$, and
$A_{\rm RD}$ is the ringdown amplitude:
\be
A_{\rm RD}=\delta_i {\cal L'}[f,f_{\rm RD}(a,M),\delta_2 Q(a)]\bar{\sigma})f^{-7/6},
\ee
where the Lorentzian function is defined by ${\cal L'}(f,f_0,\bar{\sigma})\equiv \bar{\sigma}^2/[(f-f_0)^2+\bar{\sigma}/4]$,
and $Q$ is the quality factor which depends on the final BH spin $a$.
The two amplitude parts can be combined by tanh-window functions:
\be
w^{\pm}_{f_0}={1 \over 2}\left[ 1\pm{\rm tanh} \left({4(f-f_0) \over d} \right) \right],
\ee
where $d=0.005$. The transition frequency $f_0$ is determined by 
$f_0=0.98 f_{\rm RD}$ where $f_{\rm RD}$ is a ringdown frequency given in terms of $M$ and $a$.
The effective phase is calculated by a complete SPA inspiral phasing $\psi_{\rm SPA}$, a premerger phasing $\psi_{\rm PM}$ and a ringdown phasing $\psi_{\rm RD}$ as
\be
\Psi_{\rm eff}(f) = \psi_{\rm SPA}w^-_{f_1}+\psi_{\rm PM}w^+_{f_1}w^-_{f_2}+\psi_{\rm RD}w^+_{f_2},
\label{eq.effectivephase}
\ee
with $f_1=0.1f_{\rm RD}, f_2=f_{\rm RD}$ using $d=0.005$ in the window functions.
The premerger and ringdown phasing have the forms
\bea
\psi_{\rm PM}&=&{1\over \eta} (\alpha_1 f^{-5/3}+\alpha_2 f^{-1}+\alpha_3 f^{-1/3} +  \alpha_4+\alpha_5 f^{2/3}+\alpha_6 f),  \\
\psi_{\rm RD}&=&\beta_1+\beta_2 f,
\eea
where the $\alpha_k$ coefficients are inspired by the SPA phase, redefined and phenomenologically fitted to agree with the PN-NR hybrid waveforms,
while $\beta_{1,2}$ parameters are not fitted but obtained from the premerger ansatz by taking the value and slope of the phase at the transition point   $f_{\rm RD}$.
The coefficients  introduced in this model  are tabulated in table 2 of~\cite{San10}. 
PhenomC templates were examined in the parameter range $q \leq 4$, and $ M \leq 350\msun$,
in which the model fitted the original hybrid waveforms with the fitting factors better than 0.97 for Advanced LIGO.

\subsection{Comparison}

\begin{table*}[t]
\begin{center}
\caption{\label{tab.parameterrange}{Parameter ranges of the phenomenological models valid for nonspinning BBH searches with the given detector sensitivity models.}}
\small{
\begin{tabular}{c  ccc }
\br
Model          & PhenomA    &PhenomB  & PhenomC           \\
 \mr
Mass range $[\msun]$& $50  \leq M  \leq 200$ &  $M  \leq 400$&$M  \leq 350$ \\
Mass ratio range&  $ q \leq 4$&$ q \leq 10$& $ q \leq 4$\\
Detector & Initial LIGO, Virgo, Advanced LIGO & Initial LIGO&Advanced LIGO \\
\br
 \end{tabular}}
 \end{center}
\end{table*}

In table~\ref{tab.parameterrange},
we summarise the parameter ranges, in which the validity of the phenomenological models is confirmed 
with the detector sensitivity models given in the last row.
While the validity of PhenomA is tested only in high  mass region, PhenomB and PhenomC are valid in broad mass region.
PhenomB is applicable in wider range of mass ratio $q \leq 10$, while the others are valid only in the range of $q \leq 4$.

\begin{figure}[t]
\begin{center}
\includegraphics[width=10cm]{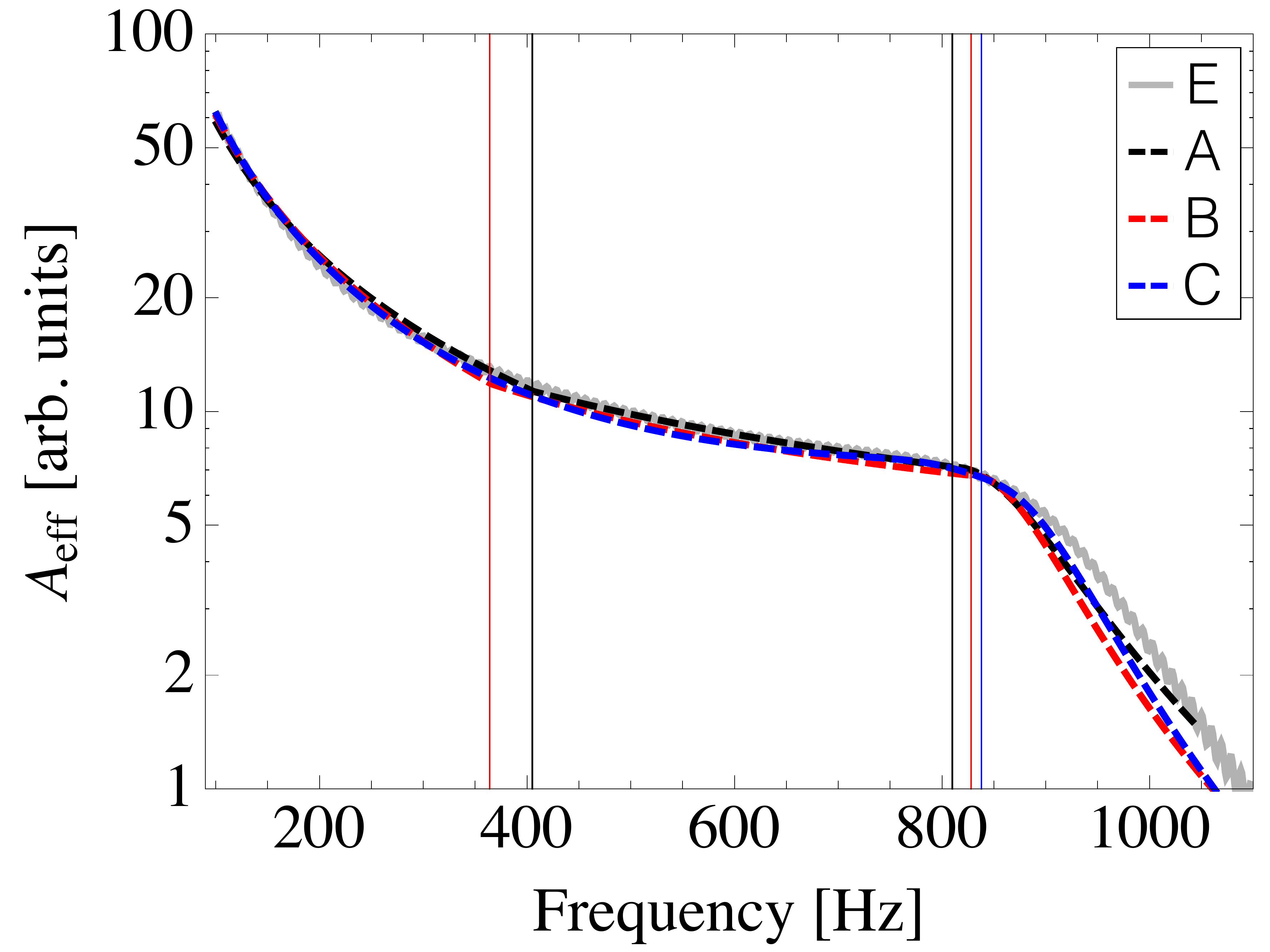}
\caption{\label{fig.ASD} Fourier-domain amplitude spectra  of EOBNRv2 (E), PhenomA (A), PhenomB (B) and PhenomC (C)  for a nonspinning BBH with masses $(10, 10) \msun$. 
The transition frequencies $\fmerg$ and $\fring$ are denoted by the black (PhenomA) and red (PhenomB) lines,
and $f_0$ (PhenomC) is denoted by the blue line.
Small oscillations in the EOBNRv2 spectrum are due to an edge effect in the fast Fourier transform, and this does not affect our analysis.}
\end{center}
\end{figure}

Figure~\ref{fig.ASD} shows the Fourier-domain amplitude spectra of EOBNRv2 and the phenomenological waveforms for a nonspinning binary with masses of $(10,10) \msun$.
The vertical lines indicate the transition frequencies that are calculated from phenomenological fits to the PN-NR hybrid waveforms.
While PhenomA and PhenomB have  two transition frequencies $\fmerg$ and $\fring$, PhenomC has one frequency $f_0$.
In this figure, we find that the phenomenological models show similar curves, and those agree well with the amplitude spectrum of the EOBNRv2 waveform.
However, the phase rather than the amplitude is a main determining factor in the match calculation, so the phase differences of the template waveforms from
the EOBNR signals will mainly affect our results of fitting factors.

\section{Validity of the phenomenological models and the effective template family}\label{sec3}

\subsection{BBH search and fitting factor}\label{sec3.1}

The matched filter is the optimal  filter for a signal of known shape in stationary Gaussian
noise. Because the waveforms emitted from coalescing binaries can be modeled reasonably well, the matched  filter can be employed in the search for BBHs.
The match between a signal ($\tilde{h}_s$) and a template ($\tilde{h}_t$) is expressed using a standard inner product defined by
\be \label{eq.match}
\langle \tilde{h}_s | \tilde{h}_t \rangle = 4 {\rm Re} \int_{\flow}^{\infty}  \frac{\tilde{h}_s(f)\tilde{h}^*_t(f)}{S_n(f)} df,
\ee
where $S_n(f)$ is a detector noise power spectral density and $\flow$ is a low frequency cutoff of the waveforms 
that depends on the shape of $S_n(f)$.
We use the zero-detuned, high-power noise power spectral density of Advanced LIGO~\cite{Lig10},
and assume $\flow=10$ Hz.

When calculating a match, we  use the normalized waveform
$\hat{h}(f) \equiv \tilde{h}(f)/ \langle \tilde{h}| \tilde{h}\rangle^{1/2}$,
then the phenomenological models for nonspinning BBHs can be given in terms of $M, \eta, t_c$ and $\phi_c$ for the match calculations.
On the other hand, the match can be easily maximized over $t_c$ and $\phi_c$~\cite{All12}, and
the two mass parameters $M$ and $\eta$ are main parameters in data analysis for nonspinning binary systems. 
Note that we use the chirp mass $M_c=M \eta^{3/5}$ instead of $M$  in this work. 
Therefore, the overlap is defined  by maximizing the match over $t_c$ and $\phi_c$ as
\be\label{eq.overlap}
P =  \max_{t_c,\phi_c}{\langle \tilde{h}_s | \tilde{h}_t \rangle   \over  \sqrt{\langle \tilde{h}_s | \tilde{h}_s \rangle \langle \tilde{h}_t | \tilde{h}_t \rangle }}.
\ee

Finally, the fitting factor (FF) is the best-match between two waveforms maximized over all possible parameters~\cite{Apo95}.
In this work, thus, FF is obtained by maximizing the overlap over $M_c$ and $\eta$ as
\be
{\rm FF} = \max_{M_c, \eta} P(M_c,\eta).
\label{eq.FF}
\ee 
For data analysis purposes, FF is used to evaluate
the search efficiency.
The gravitational wave searches use a bank of template waveforms
constructed for the corresponding mass range~\cite{Apo95,Sat91,Bal96,Owe96}.
Typically, a template bank requires that the total mismatch (i.e. 1-FF) between the templates and
signals does not exceed $3\%$~\cite{Aba12,Aas13} including the effect of discreteness of the template spacing.
In this work, in order to avoid the effect of discreteness,
we use sufficiently fine spacings in the $(M_c$-$\eta)$ plane\footnote{To obtain FF for one signal, for example, we repeat a grid search near the signal varying the search area and the template spacings until we can estimate a size of the overlap contour $\hat{P}=0.995$ roughly, where $\hat{P}$ is an overlap weighted by the maximum overlap value in that contour, and finally we find FF by performing a  $51 \times 51$ grid search in the region $\hat{P}>0.995$.}.
The detection rate is proportional to  the cube of the SNR, and
the SNR is calculated by $\rho=\langle \tilde{h}_s | \tilde{h}_s \rangle^{1/2}{\rm FF}$.
Thus,  a ${\rm FF}=0.97$  corresponds to a loss of detection rates of $\sim 10\%$ .

\subsection{FFs of the phenomenological template models}\label{sec.3.2}

\begin{figure}[t]
\begin{center}
\includegraphics[width=\columnwidth]{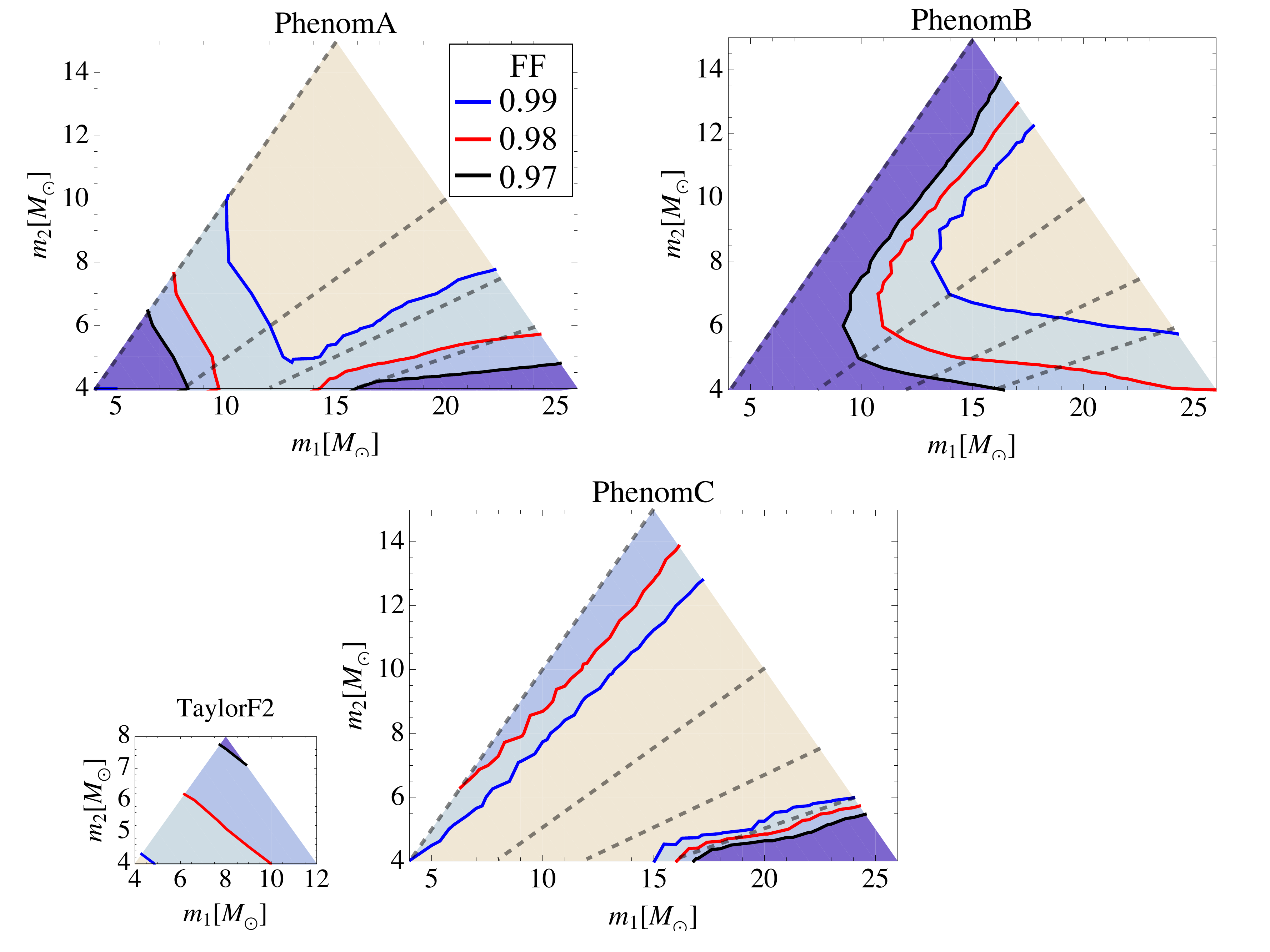}
\caption{\label{fig.ff}Fitting factors of  the  template models  to EOBNRv2 signals for nonspinning BBHs.
Darker shading indicates lower fitting factors.
Dashed lines indicate constant mass ratios, $m_1/m_2=1, 2, 3, 4$ from top to bottom.
For comparison, we also present the result of TaylorF2 inspiral template model.
Note different mass ranges between  TaylorF2 and the phenomenological models.}
\end{center}
\end{figure}

Since EOBNRv2 can model nonspinning BBH waveforms much more accurately compared to the phenomenological models,
we employ EOBNRv2 as our complete signal model 
and the phenomenological models as template models.
As described in Eq.~(\ref{eq.FF}), we calculate FF by exploring a two-dimensional overlap surface finely spaced in the $(M_c$-$\eta)$ plane.
In figure~\ref{fig.ff}, we summarise FFs of the phenomenological template models
for nonspinning BBHs with masses $m_{1,2} \geq 4\msun$ and $M\leq 30\msun$.
The blue, red and black contours correspond to ${\rm FF}=0.99, 0.98$ and 0.97, respectively.
This result shows the valid criteria of the phenomenological template models for nonspinning BBH searches 
at low mass region.
Firstly, although PhenomA has been tested only in the high mass region $M \geq 50\msun$,
our result shows that this model is also efficient in the region $M>12\msun$.
PhenomA has been tested in the region $q \leq 4$, and 
FFs in our result are also sufficiently high above 0.97 in that region. 
Secondly, PhenomB has been examined  for Initial LIGO in the range of $M \leq 400\msun$, 
but we find that FFs can be smaller than 0.97  for Advanced LIGO at the masses below $\sim15\msun$.
In addition, PhenomB has been examined in the range of $q \leq 10$, but
our result shows that this model is invalid at symmetric mass region broadly.
Finally,  the  PhenomC has been found to be valid  in broad mass region up to $\sim 350 \msun$ with $q \leq 4$ for Advanced LIGO,
and our result also shows that FFs are overall better than 0.97 in the entire low mass region with $q \leq 4$.

On the other hand, (Fourier-domain) inspiral-only templates can also be valid for BBH searches if masses of the signals are sufficiently small~\cite{Far09,Buo09,Bro13,Aji08b,Cho15}. 
Brown \etal~\cite{Bro13} has investigated FFs of TaylorF2 inspiral templates 
to EOBNRv2 signals in the mass region $3\msun \leq m_1, m_2 \leq 25 \msun$ for Advanced LIGO (see, figure 3 therein).
Their result showed that FFs  tend to decrease with increasing total mass, and the trend of FFs is very simple compared to those of the phenomenological models.
This is because TaylorF2 waveforms do not have MR phases and the cutoff frequencies are determined only by the total masses as $\pi M f_{\rm cut}=6^{-3/2}$.
In figure~\ref{fig.ff}, we also present FFs of TaylorF2 template model to EOBNRv2 signals only at the masses $M\leq 16 \msun$.
We find that the trend of FFs is very similar to that of Ref.~\cite{Bro13} but our FFs are a bit better.
A main reason of this difference is
that Ref.~\cite{Bro13} used less dense templates, consequently their result included the effect of discreteness of the template spacing.
Our result show that TaylorF2 template model is valid (i.e. ${\rm FF} \geq 0.97)$ for nonspinning BBH searches if $M \leq 15\msun$,
and FFs of that can be larger than 0.99 if $M \leq 8\msun$.
However, a comparison of FFs between TaylorF2 and PhenomC shows that 
PhenomC is still more efficient than TaylorF2 even at very low masses in our mass region.


\subsection{Constructing the effective phenomenological template family}

In figure~\ref{fig.ff},  we can find that PhenomC is most efficient among the three models for nonspinning BBH searches
although that has been developed for nonprecessing spinning BBHs.
FFs of PhenomC are always better than 0.97 independently of the mass in the region $q \leq 4$.
On the other hand, for an arbitrary signal model with masses ($m_1, m_2$), we can choose one {\it optimal} model whose FF is larger than those of the other two models. The {\it optimal} fitting factor (${\cal FF}$) is defined by
\be
{\cal FF}(m_1,m_2)={\rm Max}[{\rm A}(m_1,m_2),{\rm B}(m_1,m_2),{\rm C}(m_1,m_2)],
\ee
where ${\rm \alpha}(m_1,m_2)$ corresponds to a fitting factor calculated by using a template model $\alpha$.
Then, the {\it optimal} template family can be obtained by choosing ${\cal FF}$ for all signals in our mass region.

\begin{figure}[t]
\includegraphics[width=\columnwidth]{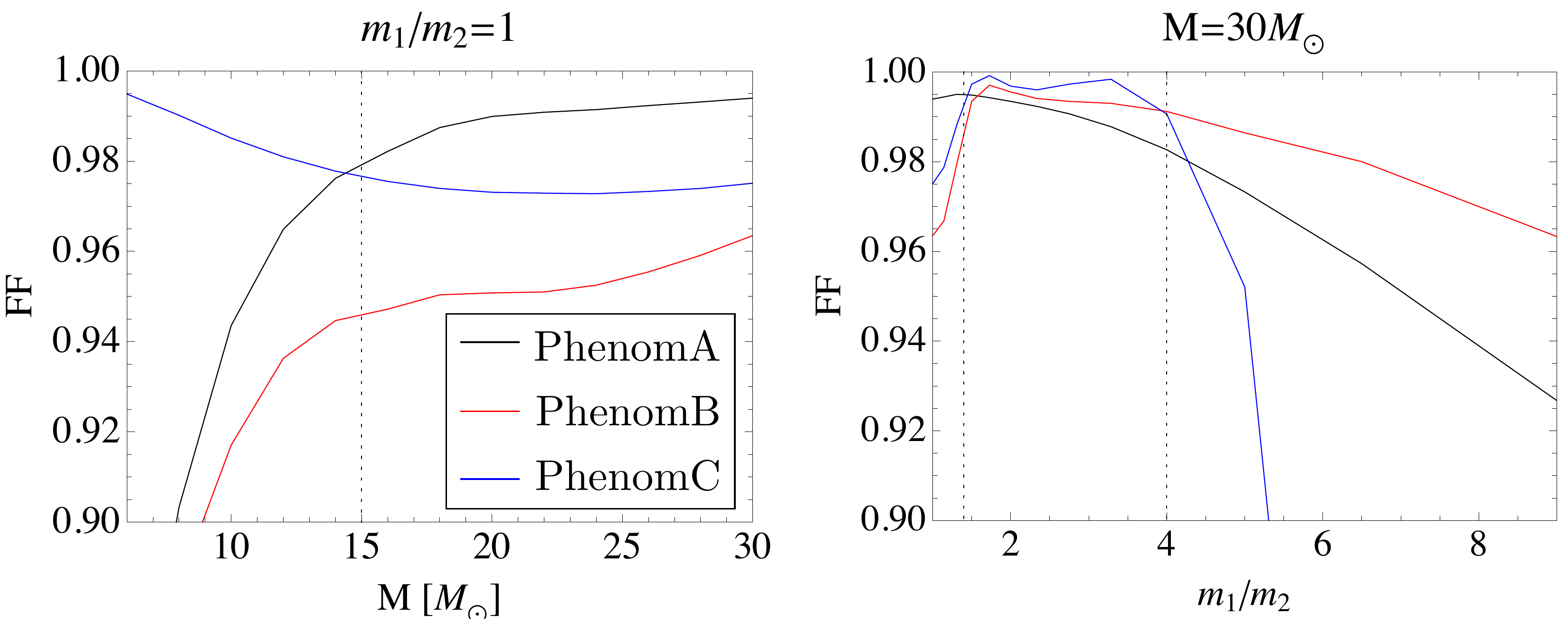}
\caption{\label{fig.ff-1d}Fitting factors calculated by changing a total mass with a fixed mass ratio $m_1 / m_2=1$ (left) and calculated by changing a mass ratio with a fixed total mass $M=30 \msun$ (right). The dotted lines indicate $M=15 \msun$ (left) and $q=1.4, 4$ (right).}
\end{figure}
\begin{figure}[t]
\begin{center}
\includegraphics[width=12cm]{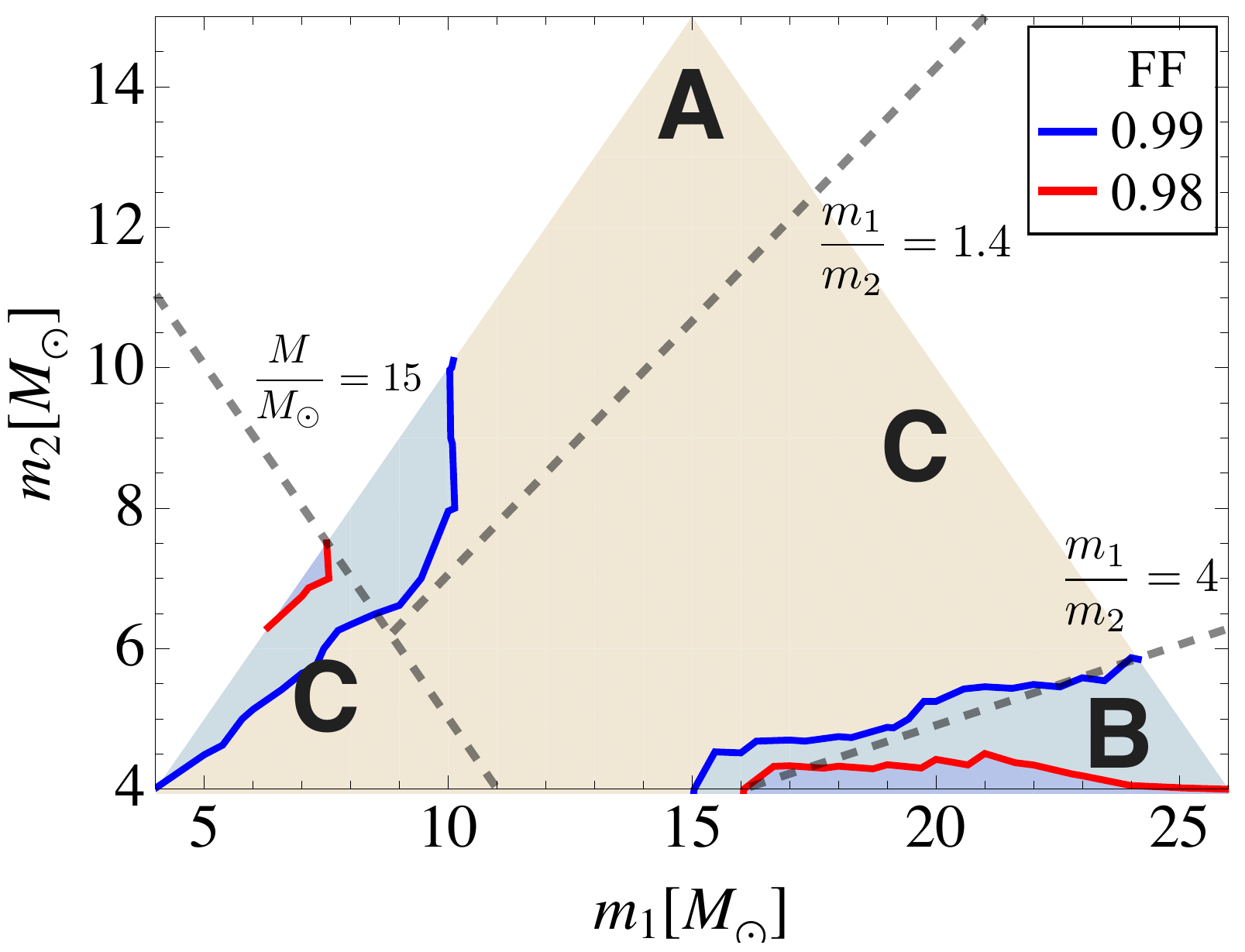}
\caption{\label{fig.ff-abc}Fitting factors of the {\it effective} template family for nonspinning BBHs calculated by using PhenomA (A), PhenomB (B) and PhenomC (C) individually in the four subregions divided by the dashed lines.}
\end{center}
\end{figure}

More simply, if we divide the  parameter region into four subregion and employ
the models in the subregions individually, an {\it effective} template family can be easily obtained.
The rule to divide the parameter region is motivated by the valid criteria of the models described in the previous subsection.
In figure~\ref{fig.ff-1d}, the left panel  shows FFs calculated by changing a total mass with a fixed mass ratio $q=1$, 
and the right panel shows FFs calculated by changing a mass ratio with a fixed total mass $M = 30\msun$.
In the left panel,  only PhenomC is valid at the masses lower than $\sim 15\msun$, while PhenomA is most efficient at $M \geq 15 \msun$.  
In the right panel, PhenomA is most efficient at $q \leq 1.4$ and PhenomB is most efficient at $q \geq 4$.
In the region $1.4 < q <4$, all models have FFs greater than 0.98, but the most efficient one is PhenomC.
Note that  $q=4$ is consistent with the boundary of the valid criteria of PhenomA and PhenomC as described in table~\ref{tab.parameterrange}.
From these features,  we divide the parameter region into four  subregions and construct an {\it effective} phenomenological template family as in figure~\ref{fig.ff-abc}.
The {\it effective} template family is composed of PhenomC waveforms in the region $M < 15 \msun$ (as shown above PhenomC is overall better than TaylorF2 in this region, so TayorF2 is unnecessary for construction of the {\it effective} template family), while in the region $M \geq 15 \msun$,
that adopts individual models   in three subregions divided by the mass ratios $q=1.4$  and 4. 
As one can see, all FFs of the {\it effective} template family are better than 0.97 and mostly greater than 0.99. 
We found that the FF contours were almost consistent with those of the {\it optimal} template family.


\section{Summary and future work}\label{sec4}

Next-generation ground-based GW detectors, such as Advanced LIGO~\cite{Aas15} and advanced Virgo~\cite{Ace15}, will start observations in coming years.
Although astrophysical BHs are likely to have spins, nonspinning  binaries may be the first search targets 
because nonspinning searches are computationally much cheaper than spinning searches. 
In addition, the Fourier-domain template models should also be used for fast matched filtering.
In this context, we investigated the validity of the existing phenomenological template models
for nonspinning BBH searches at low masses with Advanced LIGO sensitivity
assuming that EOBNRv2 can model the signal waveforms exactly.
Although PhenomB and PhenomC have been developed for spinning systems,
we also applied these models to nonspinning searches by choosing zero spins.
We found that PhenomA is valid  for low mass BBH searches in the region $M\geq 12 \msun$ and $q\leq 4$,
and PhenomB is valid in the region  $M\geq 15 \msun$ but invalid in broad asymmetric mass region.
PhenomC is most efficient among the three models, with which FFs  are   always better than 0.97 in our mass region except highly asymmetric region.
Specifically, we proposed an effective phenomenological template family
that could be easily  constructed by employing the phenomenological models in the four subregions individually.
The effective template family gives FFs better than 0.97 in our entire mass region
and mostly greater than 0.99.
The effective template family can directly applied to the search analysis, 
thus that will allow us to conduct very efficient and fast analysis for the low mass, nonspinning BBH searches with forthcoming ground-based detectors.

For spinning BBHs, PhenomB and PhenomC have been constructed by using nonprecessing PN-NR hybrid waveforms with several spin values~\cite{Aji11,San10}.
Recently, P\"{u}rrer~\cite{Pur14} has also built a Fourier-domain reduced order model using time-domain nonprecessing spinning EOBNR waveforms 
(the state-of-art version of this model is SEOBNRv2ROM, which is  implemented in the LAL). 
Using these three models, our study  can be extended to nonprecessing spinning systems,
and the feasibility of a three-dimensional effective template family can also be investigated
when an exact fiducial model for these systems is available.
On the other hand, although we only considered the phenomenological models in this work, 
our approach on determining FFs can easily be applied to the SEOBNRv2ROM template model assuming zero spins,
and our effective template family can be further improved by including the result of that model.
We will compare our  effective template family described in this work to the result of SEOBNRv2ROM, and 
investigate how much the effective template family can be improved by taking into account that model.

The necessity of using full IMR waveforms is more pronounced in high mass region
because a portion of merger-ringdown phase can significantly impact the total SNR (e.g., see figure~9 in \cite{Aji09}).
The phenomenological models are generally more efficient for high mass BBH searches.
This will also be studied in detail in a future work.

%

\ack{This work used the computing resources at the KISTI Global Science Experimental Data Hub Center (GSDC). }

%
%
%


\section*{References}
\begin{thebibliography}{9}
\bibitem{Aas15} Aasi J \etal (LIGO Scientific Collaboration) 2015 \cqg {\bf 32}, 074001 
\bibitem{Ace15} Acernese F \etal 2015 \cqg {\bf 32}, 024001 
\bibitem{Aba12}  Abadie J \etal (LIGO Collaboration, Virgo Collaboration) 2012 {\prd} {\bf 85}, 082002 

\bibitem{Far09}  Farr B, Fairhurst S and Sathyaprakash B S 2009  \cqg {\bf 26}, 114009 
\bibitem{Bro13}  Brown D A,  Kumar P and  Nitz A H 2013 {\prd} {\bf 87}, 082004 
\bibitem{Buo09}  Buonanno A,  Iyer B R, Ochsner E,  Pan Y and  Sathyaprakash B S 2009  {\prd} {\bf 80}, 084043 
\bibitem{Aji08b}  Ajith P 2008 \cqg {\bf 25}, 114033 
\bibitem{Cho15} Cho H. -S. 2015 arXiv:1502.04399

\bibitem{Aji07a}  Ajith P \etal 2007 \cqg {\bf 24}, S689 
\bibitem{Aji08a}  Ajith P \etal 2008 {\prd} {\bf77}, 104017 

\bibitem{Aji11}  Ajith P \etal 2011 {\prl} {\bf106}, 241101 
\bibitem{San10}  Santamaria L \etal 2010 {\prd} {\bf82}, 064016 
\bibitem{Han13}  Hannam M \etal 2014 {\prl} {\bf113}, 151101 
\bibitem{Buo99}  Buonanno A and Damour T 1999  {\prd} {\bf 59}, 084006 
\bibitem{Buo00}  Buonanno A and Damour T 2000  {\prd} {\bf 62}, 064015 
\bibitem{Buo07}  Buonanno A \etal 2007  {\prd} {\bf 76}, 104049 
\bibitem{Aba11}  Abadie J \etal (LIGO Collaboration, Virgo Collaboration) 2011 {\prd} {\bf 83}, 122005 
\bibitem{Pan11} Pan Y \etal 2011 {\prd} {\bf 84}, 124052 
\bibitem{Dam13}  Damour T, Nagar A and Bernuzzi S 2013  {\prd} {\bf 87}, 084035 
\bibitem{Tar12} Taracchini A, \etal 2012 {\prd} {\bf 86}, 024011
\bibitem{Tar14} Taracchini A, \etal 2014 {\prd} {\bf 89}, 061502
\bibitem{SXS} Mrou\'{e} A H \etal 2013 {\prl} {\bf 111}, 241104 Mrou\'{e} A H and Pfeiffer H P 2012 arXiv:1210.2958 
Hemberger D A \etal 2013 {\prd} {\bf 88}, 064014 Hemberger D A \etal 2013 \cqg {\bf 30}, 115001
\bibitem{Pan14a} Pan Y \etal 2014 {\prd} {\bf 89}, 084006 
\bibitem{lal} https://www.lsc-group.phys.uwm.edu/daswg/projects/lal/nightly/docs/html/




\bibitem{Pan14b} Pan Y \etal 2014  {\prd} {\bf 89}, 061501(R)
\bibitem{Szi15} Szil\'agyi B \etal 2015 {\prl} {\bf 115}, 031102





\bibitem{Lig10} {\it ``Advanced LIGO anticipated sensitivity curves"}, https://dcc.ligo.org/LIGO-T0900288/public
\bibitem{All12}  Allen B, Anderson W G,  Brady P R, Brown D A and  Creighton J D E 2012  {\prd} {\bf 85}, 122006 

\bibitem{Apo95}  Apostolatos T A 1995 {\prd} {\bf 52}, 605 
\bibitem{Sat91}  Sathyaprakash B S and  Dhurandhar S V 1991 {\prd} {\bf 44}, 3819 .
\bibitem{Bal96}  Balasubramanian R,  Sathyaprakash B S and  Dhurandhar S V 1996 {\prd} {\bf 53}, 3033 
\bibitem{Owe96}  Owen B J 1996 {\prd} {\bf 53}, 6749 

\bibitem{Aas13}    Aasi J \etal (LIGO Scientific Collaboration, Virgo Collaboration) 2013 {\prd} {\bf 87}, 022002 

\bibitem{Pur14}  P\"{u}rrer M 2014 \cqg {\bf 31}, 195010 

\bibitem{Aji09}  Ajith P and S. Bose S 2009 {\prd} {\bf79}, 084032 



\end{thebibliography}
\end{document}